\documentclass[aps,showpacs,twocolumn]{revtex4-1}
\usepackage{graphicx}
\usepackage{amsmath}
\usepackage{amssymb}
\usepackage{minibox}
\usepackage{microtype}
\usepackage{etoolbox}
\usepackage[usenames,dvipsnames,svgnames,table]{xcolor}

\makeatletter
\apptocmd{\thebibliography}{\global\c@NAT@ctr 10\relax}{}{}
\makeatother

\begin{document}

\title{Direct observation of coherent inter-orbital spin-exchange dynamics}

\author{G. Cappellini$^{1,3}$, M. Mancini$^{2,3}$, G. Pagano$^{4,3}$, P. Lombardi$^1$, L. Livi$^2$, M. Siciliani de Cumis$^{3}$,\\
P. Cancio$^{3,1,2}$, M. Pizzocaro$^5$, D. Calonico$^5$, F. Levi$^5$, C. Sias$^{3,1}$, J. Catani$^{3,1}$, M. Inguscio$^{5,2,1}$, L. Fallani$^{2,1,3}$}

\affiliation{
$^1$\minibox{LENS European Laboratory for Nonlinear Spectroscopy, Sesto Fiorentino, Italy}\\
$^2$\minibox{Department of Physics and Astronomy, University of Florence, Italy}\\
$^3$\minibox{INO-CNR Istituto Nazionale di Ottica del CNR, Sezione di Sesto Fiorentino, Italy}\\
$^4$\minibox{Scuola Normale Superiore di Pisa, Italy}\\
$^5$\minibox{INRIM Istituto Nazionale di Ricerca Metrologica, Torino, Italy}
}

\begin{abstract}
We report on the first direct observation of fast spin-exchange coherent oscillations between different long-lived electronic
orbitals of ultracold $^{173}$Yb fermions. We measure, in a model-independent way, the strength of the exchange interaction driving this coherent process. This observation allows us to retrieve important information on the inter-orbital collisional properties of $^{173}$Yb atoms and paves the way to novel quantum simulations of paradigmatic models of two-orbital quantum magnetism.
\end{abstract}


\maketitle

Alkaline-earth-like (AEL) atoms are providing a new valuable experimental platform for advancing the possibilities of quantum simulation with ultracold gases \cite{inguscio2013}. For instance, the purely nuclear spin of ground state AEL fermionic isotopes results in the independence of the atom-atom scattering properties from the nuclear spin projection. This feature has enabled the investigation of  multi-component $^{173}$Yb fermions with SU(N) interaction symmetry both in optical lattices \cite{taie2012} and in one-dimensional quantum wires \cite{pagano2014}. In addition to their nuclear spin, AEL atoms offer experimental access to supplementary degrees of freedom, in particular to a long-lived electronically-excited state $|e\rangle = |^3$P$_0\rangle$ which can be coherently populated from the ground state $|g\rangle = |^1$S$_0\rangle$ by optical excitation on an ultranarrow clock transition. The possibility of coherently manipulating both the orbital and the spin degree of freedom has recently been envisioned to grant the realization of paradigmatic models of two-orbital magnetism, like the Kondo model \cite{gorshkov2010}. In this context, the two electronic states $|g\rangle$ and $|e\rangle$ play the role of two different orbitals.

\begin{figure}[t!]
\begin{center}
\includegraphics[width=0.95\columnwidth]{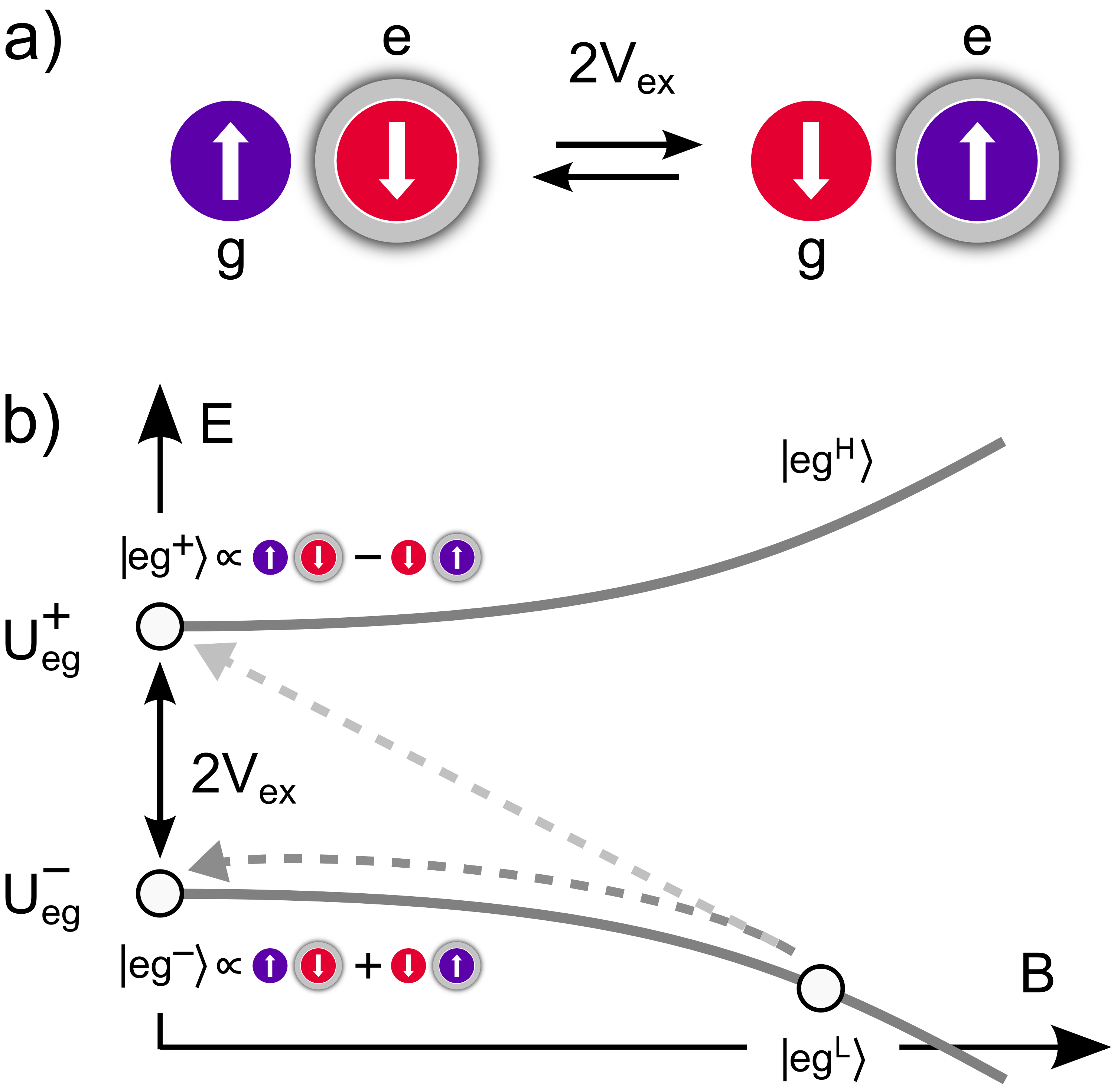}
\end{center}
\caption{Two-orbital spin-exchange interaction in AEL atoms. a) One atom in the ground state $|g\rangle$ and one atom in the long-lived electronic state $|e\rangle$ periodically ``exchange'' their nuclear spins because of the different interaction energy in the spin-singlet $|eg^+\rangle$ and spin-triplet $|eg^-\rangle$ two-particle states (note that in the graphical notation the two-particle exchange symmetry is implicit \cite{nota1}). b) Dependence of the two-particle energy on the magnetic field $B$. The spin dynamics is initiated by exciting the two atoms to the $|eg^{L}\rangle$ state at finite $B$ and then quenching the magnetic field to zero in order to create a superposition of the $|eg^+\rangle$ and $|eg^-\rangle$ states (dashed arrows).}
\end{figure}

Recent experiments have investigated the SU(N) symmetry in $|g\rangle$-$|e\rangle$ ultracold collisions of two-electron atoms \cite{zhang2014} and reported on first signatures of spin-exchange interactions between atoms in the two electronic states \cite{scazza2014}. Spin-exchange interactions arise from the difference in the spin-singlet and spin-triplet potential curves in the scattering of one $|g\rangle$ and one $|e\rangle$ atom. Let us assume that the two interacting atoms are in different nuclear spin states $|\uparrow\rangle$ and $|\downarrow\rangle$ (where the arrows are placeholders for two arbitrary nuclear spin states) and that they share the same spatial wavefunction. At zero magnetic field the degeneracy of the configurations $|g \uparrow,e \downarrow\rangle $ and $|g \downarrow,e \uparrow\rangle$, which are associated to a well-defined spin in each orbital \cite{nota1}, is lifted by the atom-atom interaction and the eigenstates are the orbital-symmetric (spin-singlet) $|eg^+\rangle$ and the orbital-antisymmetric (spin-triplet) $|eg^-\rangle$ states \cite{gorshkov2010}
\begin{equation}
|eg^\pm\rangle = \frac{1}{\sqrt{2}} \left( |g \uparrow,e \downarrow\rangle \mp |g \downarrow,e \uparrow\rangle \right)\; .
\label{eq:egpm}
\end{equation}
Owing to the different atom-atom scattering properties, these two states have different interaction energies $U_{eg}^\pm$, as sketched in Fig. 1. Preparing the two atoms in the initial state $ |\psi_0\rangle = |g \uparrow,e \downarrow\rangle = \frac{1}{\sqrt{2}} \left[ |eg^+\rangle + |eg^-\rangle \right] $ would result in a spin-exchange dynamics in which the spins of the $|g \rangle$ and $|e \rangle$ atoms are periodically flipped at a frequency  $2V_{ex}/h=|U_{eg}^- -U_{eg}^+|/h$, with probability of finding a ground-state atom in the $|g \uparrow\rangle$ state being given by
\begin{equation}
P(|g \uparrow\rangle)(t) = \frac{1}{2}\left[ 1+ \cos \left( \frac{2V_{ex}}{\hbar} t \right) \right] \; .
\label{eq:pgup}
\end{equation}
Recent measurements have suggested that in $^{173}$Yb the scattering lengths associated to the spin-triplet and spin-singlet scattering are quite different \cite{scazza2014}, resulting in a large inter-orbital spin-exchange interaction energy $V_{ex}$. However, spin oscillations induced by such interaction have not been observed, and no demonstration of the coherence of this exchange process has been shown. Here we report on the first, time-resolved observation of inter-orbital spin oscillations. This measurement clearly demonstrates the coherent nature of the exchange interaction, which is fundamentally important for its applications in quantum simulation. By measuring the oscillation frequency we determine the interaction strength $V_{ex}$ in a model-independent way, finding it to be much larger than both the Fermi energy $E_F=k_B T_F$ and $k_BT$ (where $k_B$, $T_F$ and $T$ are the Boltzmann constant, the Fermi and the gas temperature, respectively). Moreover, our measurements allow us to determine the scattering length associated with the orbital-symmetric scattering potential.

The experiment is performed on quantum degenerate Fermi gases of $^{173}$Yb in a balanced mixture of two different states out of the $I=5/2$ nuclear spin manifold, $|m_I=+5/2\rangle \equiv |\uparrow \rangle$ and $|m_I=-5/2\rangle \equiv |\downarrow \rangle$. The atoms, at an initial temperature $T \simeq 0.15 \, T_F \simeq 25$ nK, are trapped in a 3D optical lattice, with a variable depth up to $s=40$, where $s$ measures the lattice depth in units of the recoil energy $E_\mathrm{R}=h^2 /2m\lambda_L^2$, $\lambda_L$ and $m$ being the lattice wavelength and atomic mass, respectively. In our experimental conditions (see Supplemental Material \cite{supplemental}), the site occupancy in the center of the trap is $n\simeq$ 1 for each spin state. The long-lived $|e\rangle$ state is populated by exciting the  $^1 S_0$ $\rightarrow$ $^3 P_0$ intercombination transition with linearly-polarized light coming from a $\lambda=578$ nm ultranarrow laser stabilized to an ULE (Ultra Low Expansion) glass optical resonator with a closed-loop linewidth below $10$ Hz \cite{pizzocaro2012}. The lattice is operating at the magic wavelength $\lambda_L=759.35$ nm, which is not shifting the  $^1S_0$~$\rightarrow$~$^3P_0$ transition frequency \cite{barber2008}.

\begin{figure}[t!]
\begin{center}
\includegraphics[width=\columnwidth]{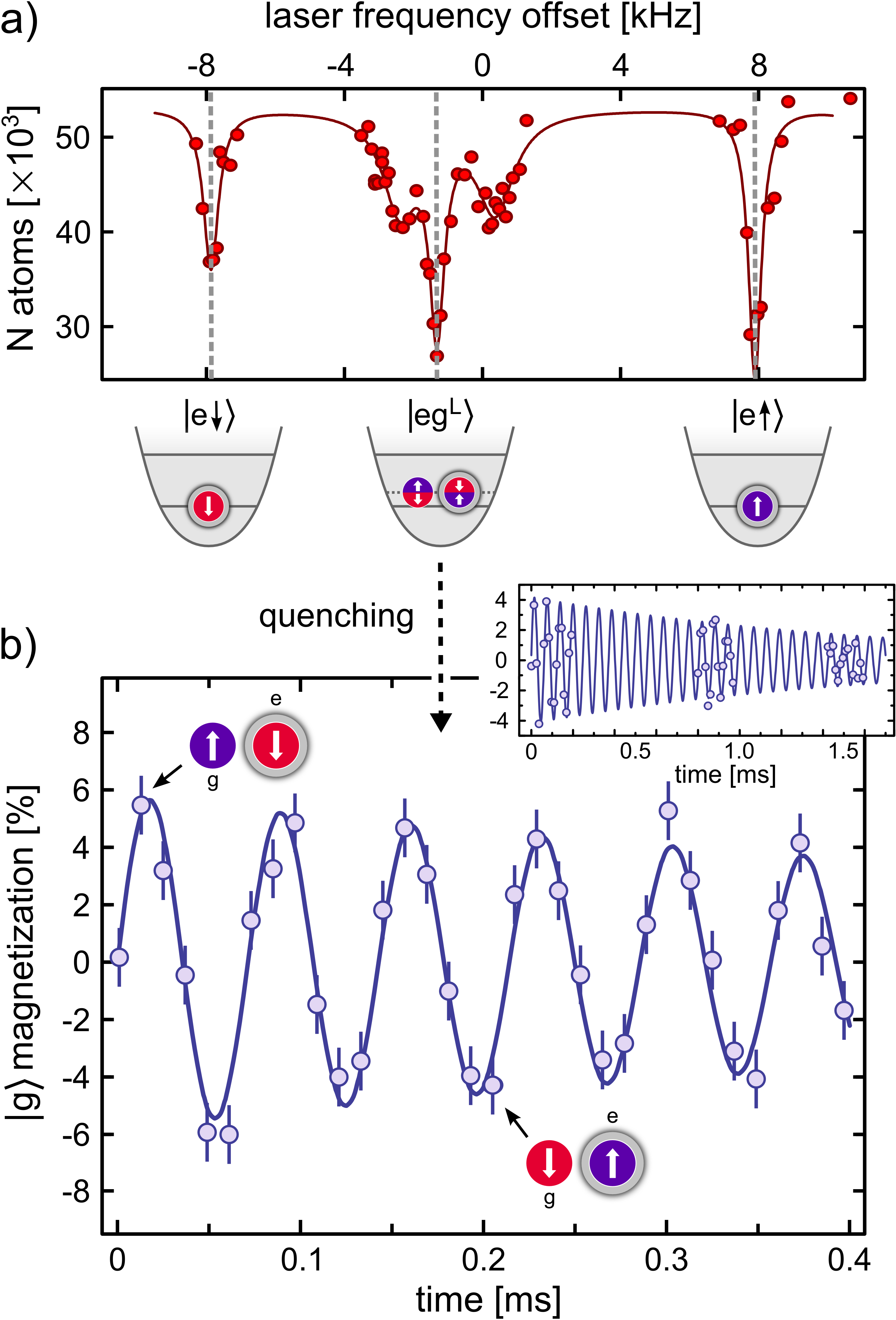}
\end{center}
\caption{a) Spectrum of the $\lambda=578$ nm clock transition for the excitation of a two-spin mixture of $^{173}$Yb atoms trapped in a 3D lattice. The vertical axis shows the number of residual $|g\rangle$ atoms after the excitation, while the horizontal axis shows the offset with respect to the clock transition frequency. The labels below the plot indicate the most prominent features of the spectrum. The dependence of the peak centers on the magnetic field $B$ allows us to attribute them to the excitation of one atom in either singly-occupied sites ($|e \downarrow \rangle$ and $| e \uparrow \rangle$) or in doubly-occupied sites ($|eg^{L}\rangle$) (see Ref. \cite{scazza2014} for the assignment of the other peaks). b) Time-resolved detection of spin-exchange oscillations. The points show the difference in fractional population between $|g \uparrow \rangle$ and $|g \downarrow \rangle$ atoms. The data shown in figure have been taken at a lattice depth $s=30.8$ after quenching the magnetic field from 60 G to a bias field of $3.5$ G. The points have been offset by a constant value ($\simeq 5\%$) to take into account a slight unbalance of the spin mixture resulting from an imperfect preparation of the initial state (which also leads to the asymmetry of the $|e \downarrow \rangle$, $| e \uparrow \rangle$ peaks in panel a). The points are averages over 5 repeated measurements and the line is the result of a fit with a damped sinusoidal function (a global error bar based on the fit residuals has been assigned to the points). The inset shows a different dataset taken at $s=35$ with oscillations extending to longer times.}
\end{figure}

A typical spectrum for a long excitation time ($\simeq 100$ ms) is reported in Fig. 2a, showing the presence of several peaks associated both to the excitation of singly- and doubly-occupied sites. We are able to spectroscopically distinguish the different peaks and address only the doubly-occupied sites by adding a static, uniform magnetic field $B$ (which was set to 28 G for the data shown in the figure). Due to the Zeeman shift, at a finite $B$ the eigenstates of the system become an admixture of spin-singlet and spin-triplet states $|eg^{L}\rangle$=$\alpha |eg^-\rangle$+$\beta |eg^+\rangle$ (\,$|eg^{H}\rangle$=$\beta^* |eg^-\rangle$$-$$\alpha^* |eg^+\rangle$\,), with $|\alpha|^2$=$|\beta|^2$=$1/2$ for infinitely large magnetic fields \cite{scazza2014} (see also Fig. 1). Note that at $B=0$ the ground state $|gg\rangle$ is coupled only to the $|eg^{L}\rangle = |eg^{-}\rangle$ state, because of the Clebsch-Gordan coefficients determining the strength of the Rabi couplings.

In order to initiate the spin dynamics we first excite the atoms with a $\pi$-pulse resonant with the $|eg^{L}\rangle$ excitation frequency. The excitation is performed at a large lattice depth $s_{in} \geq 30$, in order to avoid tunneling of atoms during the excitation time, and at large magnetic field ($60$ G), in order to have a sizeable admixture of the spin-singlet state $|eg^+ \rangle$ into the $|eg^{L}\rangle$ state ($|\alpha|^2\simeq0.75$, $|\beta|^2\simeq0.25$). Then we rapidly decrease the magnetic field to a very low bias field (3.5 G) in a time $t_\mathrm{ramp}=25$ $\mu$s, which is fast enough to have a significant population of the $|eg^{H}\rangle \simeq |eg^+\rangle$ state by nonadiabatic Landau-Zener excitation (see dashed arrows in Fig. 1) \cite{nota2}. The creation of a superposition of $|eg^-\rangle$ and $|eg^+\rangle$ states allows us to start the spin dynamics, which is observed by detecting the fraction of ground-state atoms in the different spin states by performing optical Stern-Gerlach (OSG) detection after different evolution times \cite{sleator1992}. Figure 2b shows clear oscillations of the ground-state magnetization $\left[N(g\uparrow)-N(g\downarrow)\right]/\left[N(g\uparrow)+N(g\downarrow)\right]$, which are driven by the spin-exchange process. These oscillations, clearly visible for tens of periods (as shown in the inset), provide a clear demonstration of the coherent nature of this spin-exchange interaction. The measurement of their frequency provides a direct, model-independent determination of the interaction strength, which is $2V_{ex}=h\times (13.87 \pm 0.17)$ kHz for the data in Fig. 2b, taken at $s=30.8$ \cite{nota3}.

\begin{figure}[t!]
\begin{center}
\includegraphics[width=\columnwidth]{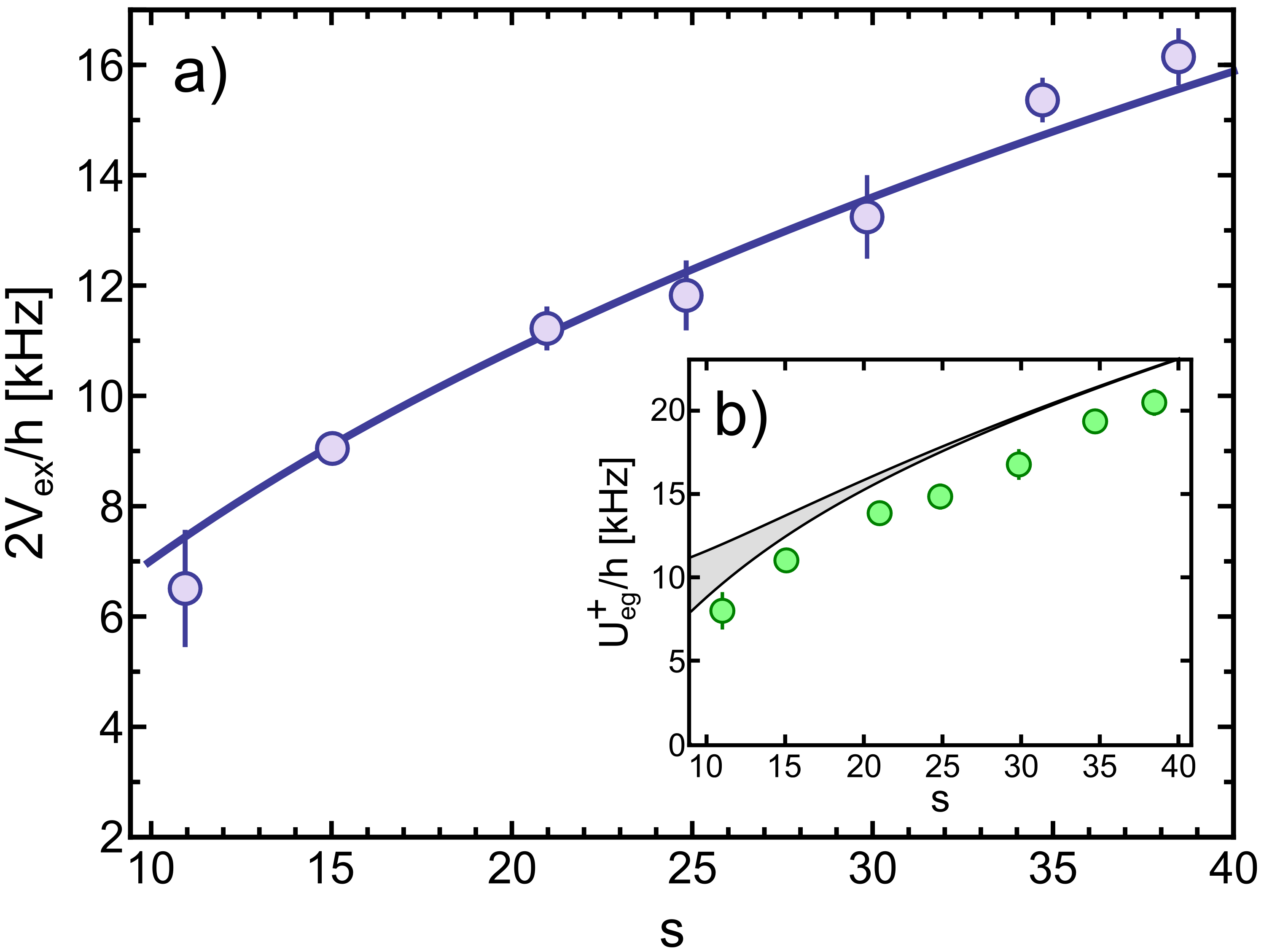}
\end{center}
\caption{a) The points show the measured spin-exchange frequency as a function of the lattice depth $s$. The data have been corrected for the small bias magnetic field $B=3.5$ G \cite{nota3} in order to show the zero-field spin-exchange frequency. Each point is the average of at least 3 different measurements and the error bar shows the statistical error. The line is a fit based on the model described in the main text.  b) The points show the interaction energy of the $|eg^+\rangle$ state, calculated as the sum of the experimentally measured $2V_{ex}$ and the $U_{eg}^-$ calculated by using $a_{eg}^-=219.5\,a_0$ \cite{scazza2014}. The shaded area shows the energy difference between ground and first excited lattice band.}
\end{figure}

We note that the relatively small amplitude of the oscillation in Fig. 2b can be ascribed to three different causes: 1) a small initial admixture of the $|eg^+ \rangle$ state in the $|eg^{L}\rangle$ state (due to excitation at a finite $B$); 2) the finite switching time of the magnetic field, which makes the projection onto the new eigenstates at low $B$ only partially diabatic; 3) the presence of singly-occupied lattice sites not participating to the spin oscillation, yet contributing to the background signal. We also have checked that these spin oscillations disappear if no laser excitation pulse is performed: collisions among $|g\rangle$ atoms can only take place in the spin-singlet channel, and the strong SU(N) interaction symmetry grants the absence of spin-changing collisions \cite{pagano2014}. We have also checked that no other nuclear spin states, different from $|\uparrow\rangle$ and $|\downarrow\rangle$, are populated during the spin-exchange dynamics.

In order to quantify the strength and the properties of the inter-orbital exchange interaction, we have measured the frequency of these spin oscillations as a function of the lattice depth $s$ and of the magnetic field $B$. 

The points in Fig. 3a show the dependence of the spin oscillation frequency $2V_{ex}/h$ on the lattice depth, clearly exhibiting a monothonic increase with $s$. In these measurements the optical excitation is performed at a lattice depth $s_{in} \geq 30$, then the optical lattice is ramped to $s$ in $\sim 700$ $\mu$s, immediately before the quench which initiates the spin dynamics. The measured values of $2V_{ex}$ are significantly large, $\approx 5$ times larger than the Hubbard interaction energy of two ground-state atoms trapped in the lattice sites, and approaches from below the energy separation between the ground and first excited band of the lattice. In this regime the usual treatment of interactions, based on the evaluation of the Hubbard onsite interaction energy with the well-known expression $U=(4\pi\hbar^2 a/m )\int |w(\mathbf{r})|^4 \, \mathrm{d}\mathbf{r}$ (where $w(\mathbf{r})$ are the single-particle Wannier functions), is expected to fail. At large interaction strength the two-particle wavefunction cannot be expressed in terms of lowest-band Wannier functions since, in the limit of infinite repulsion, the two atoms tend to spatially separate in each lattice site \cite{zurn2012} and the probability of finding them at the same position drops to zero. For a system of two particles in a harmonic potential it has been shown that, for a scattering length $a$ significantly larger than the harmonic oscillator length $a_{ho}$, the interaction energy saturates at the energy of the first excited harmonic oscillator state \cite{busch1998,kohl2006}.

In order to relate our measurements to the values of the scattering lengths $a_{eg}^\pm$ we follow a similar treatment to that adopted in Refs. \cite{deuretzbacher2008,mentink2009}, where the interaction energy for two particles in a true optical lattice potential was derived by evaluating the anharmonic corrections to the lowest-order parabolic approximation of the potential. In our analysis we express the total Hamiltonian on a basis formed by wavefunctions for the relative motion and for the center-of-mass motion of the two particles. For the former, we use the wavefunctions for interacting particles in a harmonic trap analytically derived in Ref. \cite{busch1998}; for the latter, harmonic oscillator wavefunctions are considered (see Supplemental Material \cite{supplemental} for more details). We then evaluate the anharmonic terms (up to 10$^\mathrm{th}$ order) on this basis and by numerical diagonalization of the total Hamiltonian we derive the dependence of the interaction energy in the motional ground state $U(a,s)$ as a function of the scattering length $a$ and of the lattice depth $s$. In Fig. 3a we fit the experimental data of the spin oscillation frequency vs. $s$ with the function $\left[U(a_{eg}^+,s)-U(a_{eg}^-,s)\right]/h$ (solid line), assuming the value $a_{eg}^-=219.5 \,a_0$ for the spin-triplet scattering length measured in Ref. \cite{scazza2014} (where $a_0$ is the Bohr radius). The result of the fit is a spin-singlet scattering length $a_{eg}^+=(3300 \pm 300) \, a_0$. This scattering length is remarkably large and, as shown in Fig. 3b, causes the energy of the $|eg^+\rangle$ state to almost saturate to the energy gap between the first two lattice bands (grey curve).

\begin{figure}[t!]
\begin{center}
\includegraphics[width=\columnwidth]{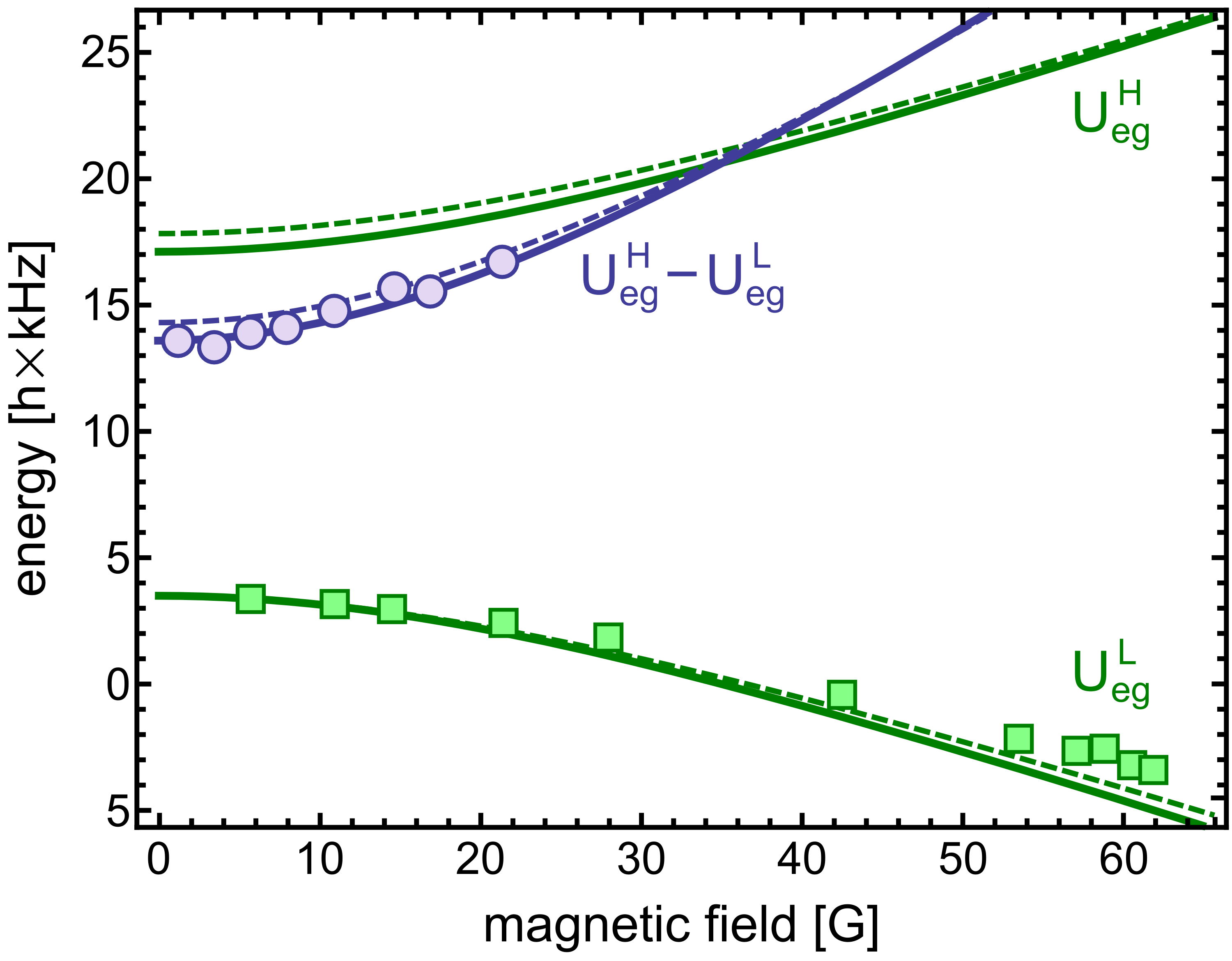}
\end{center}
\caption{Circles: measured spin-exchange frequency $(U_{eg}^H-U_{eg}^L)/h$ at $s=30$ as a function of the magnetic field. Squares: measured energy of the $|eg^L\rangle$ state derived from the spectroscopic measurements exemplified in Fig. 2a. The solid lines show the predictions of the model in Eq. (\ref{eq:matrix}) by using the $a_{eg}^+$ value derived in Fig. 3. The dashed lines show a fit of the points to the same model leaving $a_{eg}^+$ as free parameter (see main text for more details).}
\end{figure}

At a finite $B$ the spin-exchange oscillation shows a faster frequency, as the Zeeman energy increasingly contributes to the energy difference between $|eg^{L}\rangle$ and $|eg^{H}\rangle$ (see Fig. 1). The circles in Fig. 4 show the measured spin-oscillation frequency  $(U_{eg}^H-U_{eg}^L)/h$ at $s=30$ as a function of $B$, while the squares indicate the energy of the $|eg^L\rangle$ state determined by fitting the position of the peaks in the spectroscopic measurements shown in Fig. 2a. These data are compared to a simple single-band model in which the Hamiltonian of the two-atom system including interaction energy and Zeeman shift is written on the $\left\{|eg^-\rangle,|eg^+\rangle\right\}$ basis as
\begin{equation}
H = 
\begin{pmatrix}
U_{eg}^+ &
F \Delta_B \\
F \Delta_B &
U_{eg}^-
\end{pmatrix} \; ,
\label{eq:matrix}
\end{equation}
where $\Delta_B=\Delta \mu B$ is the Zeeman splitting (arising from a difference $\Delta \mu$ in the magnetic moments of the $|g\rangle$ and $|e\rangle$ states \cite{derevianko2004}) coupling the zero-field eigenstates $|eg^+\rangle$ and $|eg^-\rangle$. Differently from Ref. \cite{scazza2014}, we have included a Franck-Condon factor $F$, defined as the overlap integral
\begin{equation}
F = \iint \mathrm{d}\mathbf{r}_1 \, \mathrm{d}\mathbf{r}_2 \, \psi_{eg}^{+}(\mathbf{r}_1,\mathbf{r}_2) \psi_{eg}^-(\mathbf{r}_1,\mathbf{r}_2) \; ,
\label{eq:fc}
\end{equation}
between the wavefunctions $\psi_{eg}^\pm$ of the two atoms interacting in the two different channels. The strong repulsion in the spin-singlet channel causes indeed a strong modification of the wavefunction, resulting in an overlap integral that is significantly smaller than unity ($F\simeq 0.77$, see Supplemental Material \cite{supplemental}). By diagonalizing Eq. (\ref{eq:matrix}) we find the eigenstates $\left\{ |eg^L\rangle, |eg^H\rangle\right\}$ and the  dependence of the energies $U_{eg}^L$, $U_{eg}^H$ on the magnetic field $B$ (see also Fig. 1). The solid lines in Fig. 4 show the predictions of this model by using $a_{eg}^-=219.5 \, a_0$, $a_{eg}^+=3300\,a_0$ (from the fit in Fig. 3) and the $F$ factor calculated by using the interacting wavefunctions obtained previously. The agreement with the experimental data is quite good, showing the substantial validity of the model in Eq. (\ref{eq:matrix}) as long as the overlap factor $F$ between the interacting wavefunctions is considered. Alternatively, we have performed a simultaneous fit of the two datasets in Fig. 4 with the eigenenergies of Eq. (\ref{eq:matrix}) by expressing $U_{eg}^+$ and $F$ as functions of the free parameter $a_{eg}^+$ (obtained from the model discussed previously): the result (dashed lines) is $a_{eg}^+=(4700 \pm 700)\,a_0$, which is $\sim 2 \sigma$ away from the more precise determination coming from the fit of the data shown in Fig. 3. We note that a precise determination of $a_{eg}^+$ is complicated by the fact that, in this regime of strong interactions, the dependence of $U_{eg}^+$ on $a_{eg}^+$ is extremely weak and small effects coming e.g. from calibration uncertainties or from higher-order contributions in the theory could yield significant changes. We also note that in the presence of a tight trapping the interpretation of the results in terms of an effective scattering length should be considered \cite{bolda2002}. However, we stress that, differently from $a_{eg}^+$, our determination of $V_{ex}$ is free from any assumption or modeling and represents an accurate measurement of the spin-exchange coherent coupling in an actual experimental configuration.

The 3D lattice setting that we have used in our experiments has allowed us to study the dynamics of an isolated two-atom system in which only one atom is in the excited state, therefore significantly reducing the effects of inelastic $|e\rangle-|e\rangle$ collisions. Nevertheless, we measure a finite lifetime of the spin-exchange oscillations, on the order of $\sim 2$ ms, after which the oscillation amplitude becomes comparable with the scattering of the points (see inset in Fig. 2b). In order to investigate the origin of this damping, we have performed additional experiments in which we introduce a variable waiting time $t_{wait}$ between the laser excitation to the $|eg^L\rangle$ state and the magnetic field quench. For $t_{wait}$ as large as 30 ms (more than one order of magnitude larger than the observed damping time) we still detect high-contrast spin-exchange oscillations. This rules out the explanation of the damping in the inset of Fig. 2b in terms of either a detrimental effect of inelastic $|g\rangle-|e\rangle$ collisions in doubly-occupied sites, or a possibile collisional dephasing introduced by the tunneling of highly mobile atoms in excited lattice bands. After the exclusion of these fundamental mechanisms of decoherence, it seems highly plausible that the decay of the spin-exchange oscillations arises from technical imperfections (associated e.g. to the fast switching of the magnetic field).

In conclusion, we have observed for the first time fast, long-lived inter-orbital spin-exchange oscillations by exploiting a system of ultracold AEL fermions trapped in a 3D optical lattice. The direct observation of several periods of these oscillations has allowed us to demonstrate the coherence of the process and to measure the exchange interaction strength in an accurate, model-independent way. We note that, if compared with the spin dynamics observed in other atomic systems, arising from either small differences in the scattering lengths \cite{widera2005,krauser2012,krauser2014} or from second-order tunnelling between adjacent sites of an optical lattice \cite{trotzky2008}, the oscillation that we have measured is significantly fast. In particular, the exchange energy $V_{ex}$, on the order of $\sim h \times 10$ kHz, is much larger than either the Fermi ($k_B T_\mathrm{F}$) and the thermal ($k_\mathrm{B} T$) energies, which makes $^{173}$Yb remarkably interesting for the observation of quantum magnetism in a two-orbital system with SU(N) interaction symmetry \cite{gorshkov2010}. The direct measurement of $V_\mathrm{ex}$ has also allowed us to provide a determination of the inter-orbital spin-triplet scattering length, which exceeds the spin-singlet one by $\sim 20$ times. Besides, from a wider point of view, this strong spin-exchange interaction entangles two stable internal degrees of freedom of the atom \cite{anderlini2007}, which can be independently and coherently manipulated, opening new realistic possibilities for both quantum information processing and quantum simulation.

We would like to acknowledge N. Fabbri, M. Fattori, C. Fort and A. Simoni for useful discussions. This work has been financially supported by EU FP7 Projects SIQS (Grant 600645) and SOC-2 (Grant 263500), MIUR Project PRIN2012 AQUASIM, ERC Advanced Grant DISQUA (Grant 247371).

\renewcommand{\thefigure}{S\arabic{figure}}
 \setcounter{figure}{0}
\renewcommand{\theequation}{S.\arabic{equation}}
 \setcounter{equation}{0}
 \renewcommand{\thesection}{S.\Roman{section}}
\setcounter{section}{0}
\renewcommand{\thetable}{S\arabic{table}}
 \setcounter{table}{0}

\onecolumngrid

\newpage


\begin{center}
{\bf \large Supplemental Material for\\
``Direct observation of coherent inter-orbital spin-exchange dynamics''}\\
\bigskip
G. Cappellini, M. Mancini, G. Pagano, P. Lombardi, L. Livi, M. Siciliani de Cumis,\\
P. Cancio, M. Pizzocaro, D. Calonico, F. Levi, C. Sias, J. Catani, M. Inguscio, L. Fallani
\end{center}

\bigskip
\twocolumngrid

\section{Experimental Sequence}

\begin{figure}[t!]
\centering
\includegraphics[width=\columnwidth]{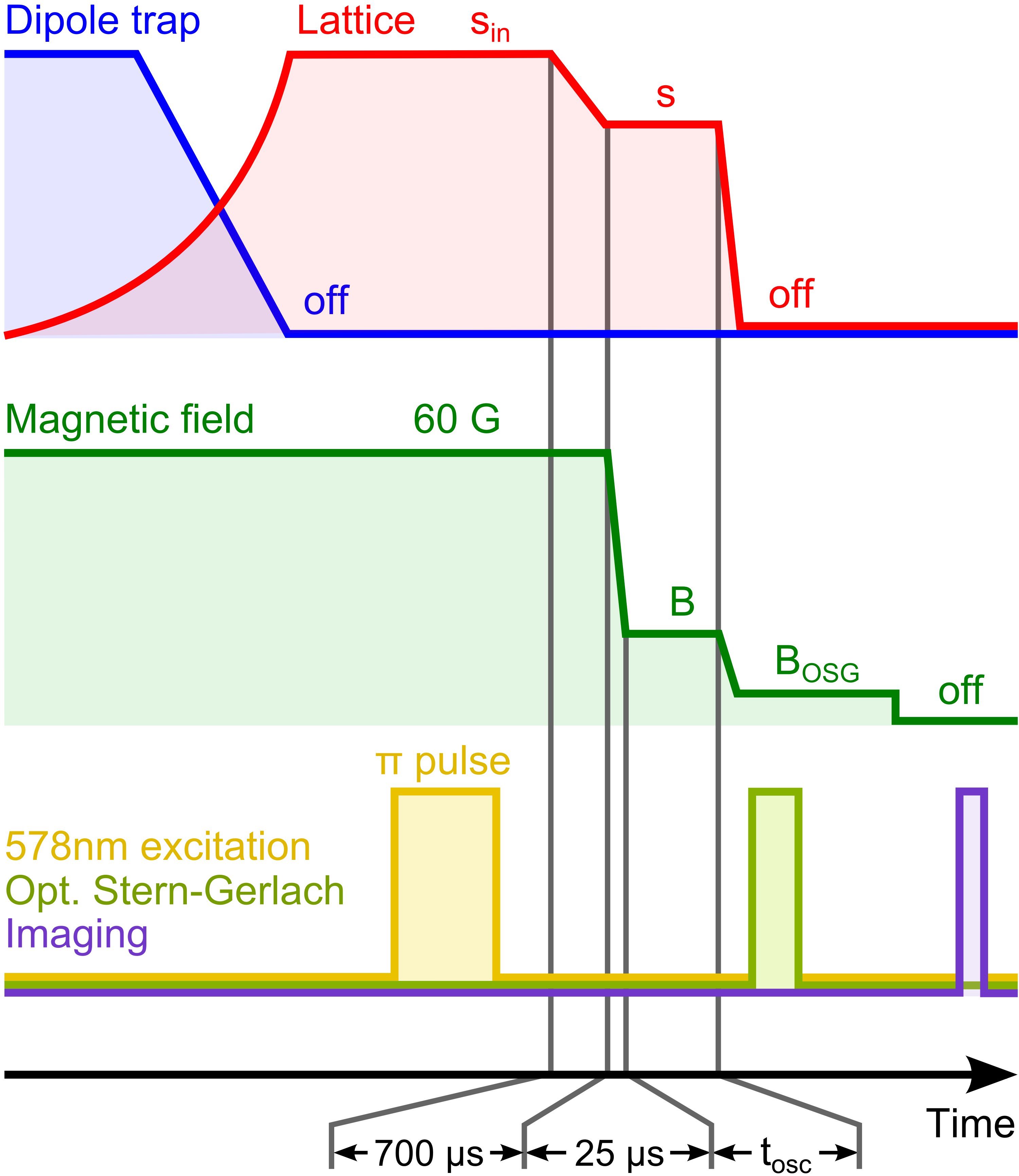}
\caption{Typical experimental sequence (see text for details).}
\label{fig:rampe}
\end{figure}

Fig. \ref{fig:rampe} shows a diagram with the time sequence of our experiments. We start with a two-component $^{173}$Yb Fermi gas ($m_{I}=\pm 5/2$), which is produced by evaporative cooling in a 1064 nm optical dipole trap until approximately $4 \times 10^{4}$ atoms are left at a temperature $T\simeq 0.15T_F \simeq 25$ nK. The atomic cloud is then adiabatically loaded in a 3D optical lattice operating at the magic wavelength $\lambda_L=759.35$ nm  and, during the same time, the optical dipole trap intensity is ramped to zero in order to let the atoms be trapped only by the lattice potential. The average filling is $0.5 \leq \overline{n}\leq 1$ atoms per lattice site and per spin component. The initial lattice depth is $s_{in}\geq 30$ (in units of the recoil energy $E_\mathrm{R}=h^2 /2m\lambda_L^2$, where $m$ is the atomic mass).  

The atoms are excited by a 578 nm $\pi$-pulse, resonant with the $|gg\rangle\rightarrow|eg^L\rangle$ transition, at a high magnetic field $B \simeq 60$ G. After the excitation pulse the lattice depth is quickly ramped to $s$ in $\approx 700$ $\mu$s and, immediately after, the magnetic field is quenched to the final $B$ value in $25$ $\mu$s, sufficiently fast to have a significant projection of the atomic state onto $|eg^{+}\rangle$. At this point the spin-exchange oscillation  $|g\uparrow,e\downarrow \rangle \leftrightarrow |g\downarrow,e\uparrow\rangle$ is started. After a variable oscillation time $t_{osc}$, the optical lattice is switched off and the populations in the $|g\uparrow\rangle$, $|g\downarrow\rangle$ states are measured after an optical Stern-Gerlach pulse, followed by a time of flight of 4.5 ms.

\section{Theoretical Model}

Here we describe the model that we have developed in order to relate the large interaction energies measured in the experiment to the values of the scattering lengths $a_{eg}^{\pm}$ describing the $s$-wave collisions of two $^{173}$Yb atoms in the $|g\rangle+|e\rangle$ channel. This model is valid also for strong interactions, when the relation between the Hubbard interaction energy and the scattering length $a$ is no longer linear, as it is in the usual expression $U_{\mathrm{Hub}}= \frac{4\pi \hbar^{2}}{m}a \int |w(\mathbf{r})|^{4}\,\mathrm{d}\mathbf{r} $, where $w(\mathbf{r})$ is the lowest-band Wannier function for a (noninteracting) atom localized at a lattice site \cite{jaksch1998sup}.

The Hamiltonian describing two atoms interacting in a lattice potential well is:
\begin{equation}
H = \frac{p_{1}^{2}}{2m}+\frac{p_{2}^{2}}{2m}+V_{lat}(\mathbf{r}_{1})+V_{lat}(\mathbf{r}_{2})+ V_{int}(\mathbf{r}_{1}-\mathbf{r}_{2}) \; , 
\end{equation}
where  $V_{lat}(\mathbf{r})= V_{0}\sum_{i= x,y,z} \sin^{2}(k r_{i})$ is the lattice potential experienced by each atom and  $V_{int}(\mathbf{r})= \frac{4\pi \hbar^{2}}{m}a \, \delta(\mathbf{r})\frac{\partial}{\partial r}r $ is the interaction potential, expressed in the form of a regularized pseudopotential \cite{busch1998}.

In order to take into account the anharmonicity of the lattice potential (which is essentially important for a quantitative comparison with the experimental data), we expand $V_{lat}(\mathbf{r})$ around the origin up to the $10^{\mathrm{th}}$ order:
\begin{equation}
V_{lat}(\mathbf{r})= V_{0}\sum_{i=x,y,z}(k^{2} r_{i}^{2}-\frac{1}{3}k^{4} r_{i}^{4}+\frac{2}{45}k^{6}r_{i}^{6}+...) \; .
\end{equation}
This order of expansion is high enough to describe properly the shape of an individual lattice well (in order to consider the effects of tunneling, which are important only at low lattice depth, the potential should be expanded to a higher order, at least to the $20^{\mathrm{th}}$, making the problem computationally much longer to solve). Introducing $\omega = 2\sqrt{s} E_{rec}/\hbar$ we can rewrite the Hamiltonian as 
\begin{multline}
H = \frac{p_{1}^{2}}{2m}+\frac{p_{2}^{2}}{2m}+\frac{1}{2}m\omega^{2}r_{1}^{2}+\frac{1}{2}m\omega^{2}r_{2}^{2} \\
+ V_{int}(\mathbf{r}_{1}-\mathbf{r}_{2})+ V_{anh}(\mathbf{r}_{1},\mathbf{r}_{2}) \; ,
\end{multline}
where $V_{anh}(\mathbf{r}_{1},\mathbf{r}_{2})$ contains the anharmonic terms coming from the expansion of the lattice potential.
By making the substitution $\mathbf{R} = \frac{\mathbf{r}_1+\mathbf{r}_2}{\sqrt{2}}$ and $\mathbf{r}=\frac{\mathbf{r}_1-\mathbf{r}_2}{\sqrt{2}}$, we can write the Hamiltonian in terms of center-of-mass $\left\{\mathbf{R},\mathbf{P} \right\}$ and relative $\left\{\mathbf{r},\mathbf{p} \right\}$ coordinates: 
\begin{multline}
H = \frac{P^{2}}{2m}+\frac{p^{2}}{2m}+\frac{1}{2}m\omega^{2}R^{2}+\frac{1}{2}m\omega^{2}r^{2}+ V_{int}(\mathbf{r})\\
+ V_{anh}(\mathbf{R},\mathbf{r}) \; .
\label{h}
\end{multline}
The harmonic+interaction part of the Hamiltonian (including all the terms in Eq. (\ref{h}) except the last one) was solved analytically by Busch et al. \cite{busch1998sup}. This work showed that the interaction energy for two atoms in the ground state of the trap saturates at the energy of the first excited vibrational state in the limit of $a \rightarrow +\infty$. For a true lattice potential, the anharmonic terms $V_{anh}(\mathbf{R},\mathbf{r})$ couple the relative and center-of-mass motion, making the problem impossibile to be solved analytically. 

In order to extend the results of Busch et al. to the case of a lattice potential well, we diagonalize numerically the full Hamiltonian in Eq. (\ref{h}) written on a basis of wavefunctions which are solutions of the harmonic problem: 
\begin{equation}
\Psi_{N,L,M}({\mathbf{R}})\;\phi_{n,l,m}({\mathbf{r}}) \; ,
\end{equation}
where $N$ ($n$) is the principal quantum number and $L,M$ ($l,m$) are the angular momentum quantum numbers for the center-of-mass (relative) motion. For the relative wavefunctions $\phi_{n,l,m}({\mathbf{r}})$ we choose the solutions of the 3D isotropic harmonic oscillator for $l\neq 0$, while for $l = 0$ we take the interacting wavefunctions derived in Ref. \cite{busch1998sup}
\begin{multline}
\phi({\mathbf{r}}) = A \exp\left(-\frac{r^2}{2 a_{ho}^2}\right)\Gamma\left( -\frac{E}{2\hbar \omega}+\frac{3}{4} \right) \\ U \left(-\frac{E}{2\hbar \omega}+\frac{3}{4},\frac{3}{2},\frac{r^2}{a_{ho}^2} \right) \;,
\end{multline}
where $U$ are the confluent hypergeometric functions, $A$ is a normalization factor, $a_{ho}=\sqrt{\hbar/m \omega}$ is the harmonic oscillator length and $E$ is the total energy, given by the solution of the equation
\begin{equation}
\sqrt{2} \frac{\Gamma\left(-\frac{E}{2\hbar \omega}+\frac{3}{4}\right)}{\Gamma\left(-\frac{E}{2\hbar \omega}+\frac{1}{4}\right)} = \frac{a_{ho}}{a} \;.
\end{equation}
For the center-of-mass wavefunctions $\Psi_{N,L,M}({\mathbf{R}})$ we always choose the solutions of the harmonic oscillator problem. We found that taking $N_{max}= n_{max}=4$ (corresponding to 196 states forming the basis) is sufficient to ensure convergence in the calculation of the ground-state energy. 

\begin{figure}[t!]
\centering
\includegraphics[width=\columnwidth]{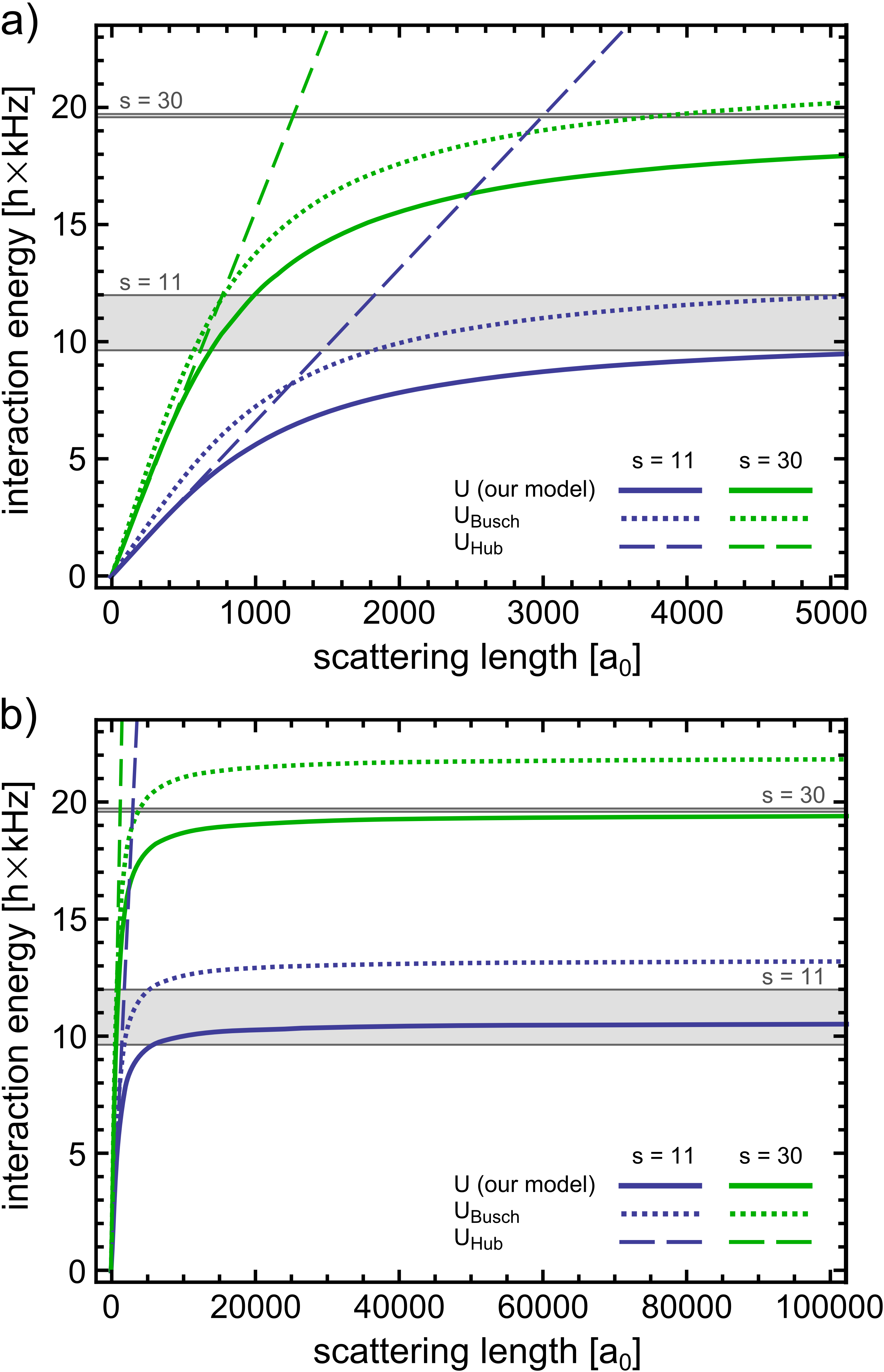}  
\caption{a) Interaction energies for two particles in a lattice site, calculated for two lattice depths $s=11$ and $s=30$ according to three different models (see text). The interaction energy $U$ calculated with our model is well approximated by the usual Hubbard relation $U_\mathrm{Hub}$ at small scattering length $a$. b) The same results are plotted up to larger values of $a$. For large $a$ the interaction energy $U$ saturates at the energy difference between the ground and the first-excited lattice band, here represented by the grey regions (the width of these regions reflects the finite width of the energy bands caused by tunnelling).}
\label{fig:uint}
\end{figure}

In Fig.\,\ref{fig:uint} we plot the results for the interaction energy (defined as the total energy minus the total energy in the noninteracting case) as a function of the scattering length $a$ for two values of the lattice depth $s=11$ and $s=30$. The curves are based on three different models: 1) our model, containing anharmonic terms and the coupling between relative and center-of-mass motion ($U$, solid lines); 2) the model of Ref. \cite{busch1998sup}, containing only the harmonic part of the potential ($U_{\mathrm{Busch}}$, dotted lines); 3) the usual expression for the interaction energy in the Hubbard model \cite{jaksch1998sup}, which takes into account the full lattice potential and depends linearly on $a$ ($U_{\mathrm{Hub}}$, dashed lines). In addition, the first band gaps for $s = 11$ and for $s = 30$ are shown. The interaction energy derived from our model saturates at the first excited band of the lattice for large values of the scattering length and, for low $a $, it is well approximated by the usual Hubbard expression $U_{\mathrm{Hub}}$. Instead, the $U_{\mathrm{Busch}}$ curves saturate at a higher energy, coincident with $\hbar \omega = 2\sqrt{s}E_{rec}$.

\begin{figure}[t!]
\centering
\includegraphics[width=\columnwidth]{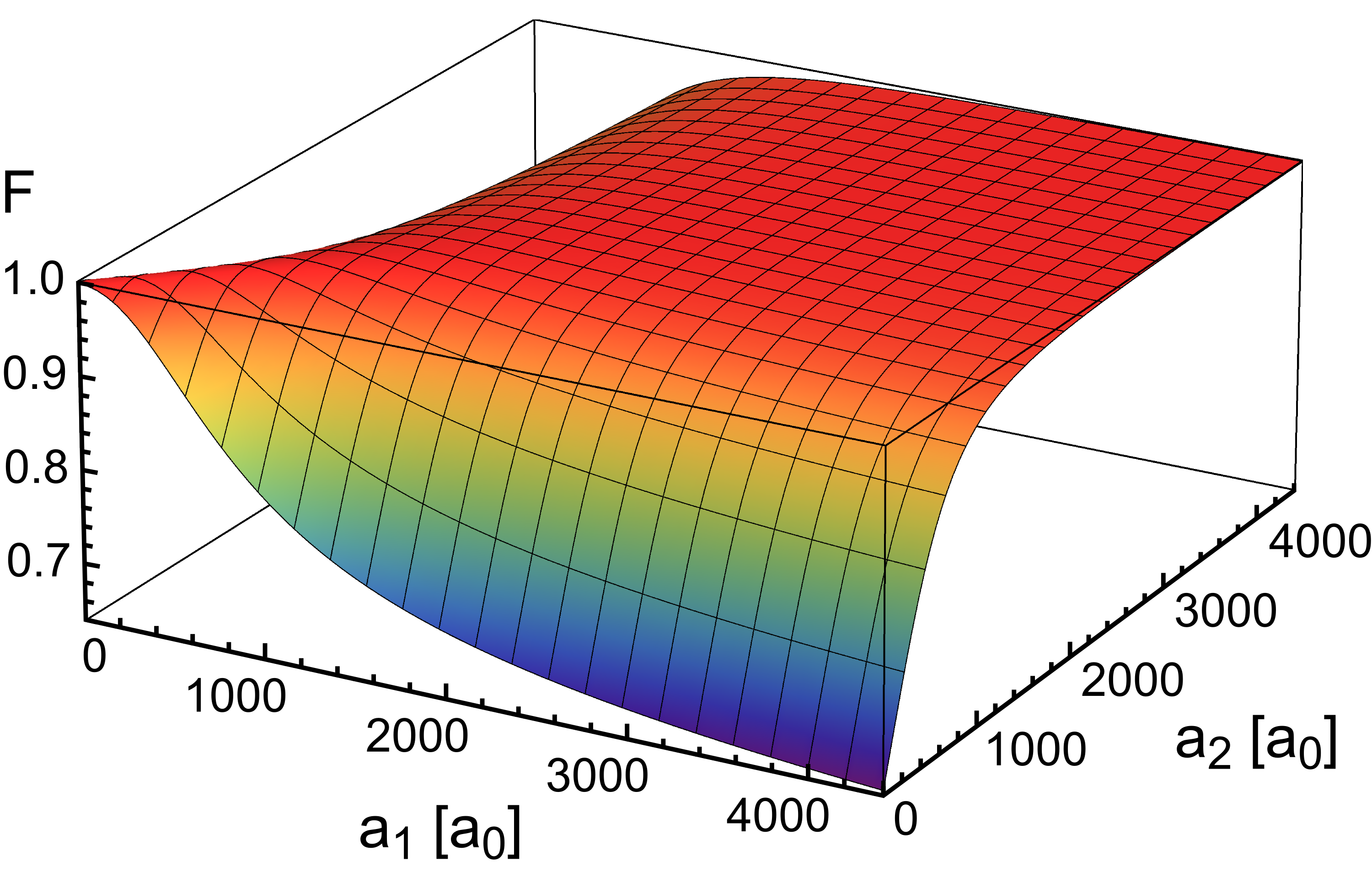}  
\caption{Franck-Condon factor $F(a_1,a_2) = \langle \psi (a_{1})\vert\psi (a_{2})\rangle$ describing the overlap of the ground-state wavefunctions for two different scattering lengths $a_1$ and $a_2$.}
\label{fig:franckcondon}
\end{figure}

By evaluating the eigenstates of the interacting system, we can compute the Franck-Condon factors $F$ that must be put in the off-diagonal elements of the matrix in Eq. (3) of the main text. The Franck-Condon factor $F$ is defined as the overlap $\langle \psi (a_{1})\vert\psi (a_{2})\rangle$ where $\psi (a)$ is an eigenstate of the Hamiltonian in Eq. (\ref{h}) with the scattering length $a$.
In Fig.\,\ref{fig:franckcondon} we plot the Franck-Condon factors between two ground states of the system for different scattering lengths.
We can see that along the diagonal (where $a_{1} = a_{2}$) the Franck-Condon factor is unity, as expected since the two states coincide, while it drops down to $\sim 0.6$ for the maximal difference in scattering lengths. 

\section{Ultranarrow 578 \lowercase{nm} laser}

The laser radiation at 578 nm used to excite the atoms to the metastable $|e\rangle$ $=$ $^3P_0$ state is produced by second-harmonic generation of the 1156 nm infrared light emitted by a quantum dot laser. Employing a bow-tie optical cavity to enhance the efficiency of the frequency doubling process, we obtain up to 50 mW of 578 nm light. A small part of this radiation is coupled into a 10 cm long ULE (Ultra-Low Expansion) glass cavity, originally employed to realize the clock laser for the Yb optical lattice clock experiment running at INRIM \cite{pizzocaro2012sup}. 

The laser frequency is locked to the ULE cavity with a 500 kHz bandwidth feedback system, and the in-loop linewidth of the laser can be estimated from the frequency noise spectrum to be below 10 Hz \cite{hall1992sup}. The ULE cavity, surrounded by a thermally-stabilized copper shield, is located in a $10^{-7}$ mbar vacuum chamber to greatly reduce its mechanical and thermal sensitivity. The whole system is placed on an antivibration platform to further reduce seismic noise, and is enclosed in an isolation box to decouple the system from the lab environment.

The long-term drift of the cavity has been characterized and is corrected excluding a residual drift on the order of 100 Hz/day. However, erratic fluctuations of some Hz/s, that we ascribe to an imperfect thermal stabilization of the ULE cavity, limit our mid-term stability and represent one of the limitations in the observation of long spin-exchange oscillations.


\begin{thebibliography}{99}
\bibitem[7]{nota1}
In the notation $|g \uparrow,e \downarrow\rangle$ the two-particle exchange symmetry is implicit. The full state has to be intended as $|g \uparrow,e \downarrow\rangle = \left(|g \uparrow \rangle_1 |e \downarrow \rangle_2 - |e \downarrow \rangle_1|g \uparrow \rangle_2 \right)/\sqrt{2} $, which is antisymmetric for the exchange of fermion 1 with fermion 2.
\bibitem[11]{nota2}
In the limit $t_\mathrm{ramp}\rightarrow 0$ the relative population of the $|eg^+\rangle$ and $|eg^-\rangle$ states would be $\beta^2/\alpha^2$.
\bibitem[13]{nota3}
We note that the finite bias magnetic field $B \simeq 3.5$ G used for the measurement yields a slightly faster oscillation frequency than $2V_{ex}/h$ (by $\sim 100$ Hz). In order to show the zero-field oscillation frequency the datapoints in Fig. 3 have been corrected by using the finite-$B$ model described later in the text.
\end{thebibliography}
\end{document}